\def\gtwid{\mathrel{\raise.3ex\hbox{$>$\kern-.75em\lower1ex\hbox{$\sim$}}}}
\def\ltwid{\mathrel{\raise.3ex\hbox{$<$\kern-.75em\lower1ex\hbox{$\sim$}}}}
\def\Box{\kern1pt\vbox{\hrule height 1.2pt\hbox{\vrule width 1.2pt\hskip 3pt
   \vbox{\vskip 6pt}\hskip 3pt\vrule width 0.6pt}\hrule height 0.6pt}\kern1pt}
\documentclass[12pt]{article}
\input{psfig.sty}
\begin{document}
\begin{titlepage}
\begin{flushright}
hep-ph/9710444 \\ UFIFT-HEP-97-22 \\ CRETE-97-13
\end{flushright}
\vspace{.4cm}
\begin{center}
\textbf{The Quantum Gravitationally Induced Stress Tensor \\
During Inflation}
\end{center}
\begin{center}
N. C. Tsamis$^{\dagger}$
\end{center}
\begin{center}
\textit{Department of Physics \\ University of Crete \\ 
Heraklion, CR-71003 GREECE}
\end{center}
\begin{center}
R. P. Woodard$^*$
\end{center}
\begin{center}
\textit{Department of Physics \\ University of Florida \\ 
Gainesville, FL 32611 USA}
\end{center}
\begin{center}
ABSTRACT
\end{center}
\hspace*{.5cm} We derive non-perturbative relations between the expectation
value of the invariant element in a homogeneous and isotropic state and
the quantum gravitationally induced pressure and energy density. By exploiting 
previously obtained bounds for the maximum possible growth of perturbative
corrections to a locally de Sitter background we show that the two loop result
dominates all higher orders. We also show that the quantum gravitational
slowing of inflation becomes non-perturbatively strong earlier than previously
expected.
\begin{flushleft}
\end{flushleft}
\vspace{.4cm}
\begin{flushleft}
$^{\dagger}$ e-mail: tsamis@physics.uch.gr \\
$^*$ e-mail: woodard@phys.ufl.edu
\end{flushleft}
\end{titlepage}

\section{Introduction}

Gauge-fixed perturbation theory is by far the simplest method for computing
quantum corrections to a classical geometry. Even when the state of interest 
is not stationary this can be done using Schwinger's formalism for expectation 
values \cite{schw,rj}. The procedure is first to compute the expectation value 
of the invariant element in the presence of the desired state:
\begin{equation}
\left\langle \psi \left\vert g_{\mu\nu}(t,{\vec x}) dx^{\mu} dx^{\nu} 
\right\vert \psi \right\rangle = {\widehat g}_{\mu\nu}(t,{\vec x}) 
dx^{\mu} dx^{\nu} \; .
\end{equation} 
One then forms ${\widehat g}_{\mu\nu}$ into gauge invariant and gauge
independent observables to infer how quantum effects distort the geometry.

Geometrically significant {\it differences} between the classical and quantum
backgrounds can be ascribed to a quantum-induced stress tensor. In pure
gravity this is defined from the deficit by which ${\widehat g}_{\mu\nu}$
fails to obey the classical Einstein equation:
\begin{equation}
8 \pi G {\widehat T}_{\mu\nu} \equiv {\widehat R}_{\mu\nu} - \frac12 {\widehat 
g}_{\mu\nu} {\widehat R} + {\widehat g}_{\mu\nu} \Lambda \; . \label{eq:stress}
\end{equation}
Here ${\widehat R}_{\mu\nu}$ and ${\widehat R}$ are the Ricci tensor and Ricci
scalar constructed from ${\widehat g}_{\mu\nu}$ and it should be noted that we 
have included a cosmological constant $\Lambda$ in Einstein's equation. Note 
also that the relation between the induced stress tensor and the quantum 
background ${\widehat g}_{\mu\nu}$ is, in principle, non-perturbative, even
though the only practical way of computing ${\widehat g}_{\mu\nu}$ is
perturbatively.

The purpose of this paper is to derive the leading late time dependence, 
{\it to all orders}, for the induced stress tensor appropriate to a recent 
calculation of the quantum gravitational back-reaction on an initially empty 
and inflating universe \cite{tw1}. That we can obtain an all-orders result
arises from the conjunction of the non-perturbative relation (\ref{eq:stress})
and explicit bounds on the maximum late time growth of perturbative 
corrections to a rather technical variant of the amputated 1-point function.
Section 2 reviews the definition of this quantity and the procedure through
which it is used to compute ${\widehat T}_{\mu\nu}$. Section 3 shows how the
perturbative bounds on the former imply an all-orders result for the latter.
We discuss the consequences of this result in Section 4. In what remains of
this Introduction we review the theoretical context of our previous work and
its physical motivation.

Because the late time behavior is dominated by ultraviolet finite, non-local
terms, we were able to use the Lagrangian of general relativity with a 
positive cosmological constant:
\begin{equation}
{\cal L} = {1 \over 16 \pi G} \left(R - 2 \Lambda\right) \sqrt{-g} + 
{\rm counterterms} \; ,
\end{equation}
absorbing ultraviolet divergences with local counterterms as required. We
worked on the manifold $T^3 \times R$ in the presence of a homogeneous and
isotropic state for which the invariant element takes the following form in
co-moving coordinates:
\begin{equation}
{\widehat g}_{\mu\nu}(t,{\vec x}) dx^{\mu} x^{\nu} = - dt^2 + \exp[2 b(t)]
d{\vec x} \cdot d{\vec x} \; . \label{eq:element}
\end{equation}
Our state is free de Sitter vacuum at $t=0$ in these coordinates, 
corresponding to the following classical background:
\begin{equation}
b_{\rm class}(t) = H t \qquad , \qquad H^2 \equiv \frac13 \Lambda \; .
\label{eq:class}
\end{equation}

The physical motivation for our work is the possibility that the cosmological
constant only appears to be unnaturally small today because it is screened
by an infrared process in quantum gravity. This process is the buildup of 
gravitational interaction energy between virtual gravitons that are pulled 
apart by the inflationary expansion of the classical background 
(\ref{eq:element}-\ref{eq:class}). The effect acts to slow inflation because 
gravity is attractive. It requires a enormous time to become significant 
because gravity is a weak interaction, even for inflation on the GUT 
scale.\footnote{One traditionally defines the ``scale of inflation'' $M$ so
that $M^4$ equals the energy density of the cosmological constant, 
$\Lambda/(8\pi G)$. Since the Planck mass is $M_P = G^{-1/2}$, the 
dimensionless coupling constant that characterizes quantum gravitational 
effects on inflation can be expressed as:
\begin{equation}
G \Lambda = 8 \pi \left({M \over M_P}\right)^4 \; .
\end{equation}
For GUT scale inflation this works out to about $G \Lambda \sim 10^{-11}$.
The comparable figure for inflation on the electroweak scale would be about
$G \Lambda \sim 10^{-67}$.}
However, inflation must eventually be ended because the effect adds coherently 
for as long as exponential expansion persists. The effect is also unique to 
gravitons. Only massless particles can give a coherent effect, and the other 
phenomenologically viable quanta of zero mass are prevented from doing so by 
conformal invariance.

Our mechanism offers a natural explanation for how inflation can have lasted 
a long time, without fine tuning and without the need for fundamental scalars.
Indeed, it results in such a long period of inflation that all energetically 
favorable phase transitions may have time to occur during this period, even if
some are subsequently reversed by re-heating. If so, the cosmological constant 
which is finally screened would be that of the true vacuum, and the evolution 
after inflation would be almost that which is usually obtained by keeping
gravity classical and fine tuning this parameter to zero.

Although perturbation theory must break down at the end of inflation, one can
use the technique to partially verify our proposal. For example, the presence
of infrared divergences in in-out matrix elements \cite{tw2} and scattering 
amplitudes \cite{tw3} invalidates the null hypothesis that inflation persists 
to asymptotically late times with only perturbatively small corrections. One 
can also use Schwinger's formalism to follow the evolution of the background 
until quantum corrections become non-perturbatively large \cite{tw1}. It was
previously believed that this occurred at the same time for all orders. The
burden of this paper is to show that in fact the two loop effect becomes 
strong at a time when all higher orders are still insignificant. Of course one
cannot extend past the breakdown of perturbation theory by using perturbation
theory, but we now have precise information about how the breakdown occurs.

\section{Perturbation Theory Revisited}

The purpose of this section is to explain the connection between the
induced stress tensor of co-moving coordinates and the quantities we
actually computed. We begin with the coordinate system of the classical 
background. For a variety of reasons, it is simplest to formulate 
perturbation theory in conformal coordinates:
\begin{equation}
-dt^2 + \exp[2 H t] \; d{\vec x} \cdot d{\vec x} = \Omega^2
\left(-du^2 + d{\vec x} \cdot d{\vec x}\right) \; ,
\end{equation}
\begin{equation}
\Omega \equiv {1 \over H u} = \exp(H t) \; .
\end{equation}
Note the temporal inversion and the fact that the onset of inflation at 
$t=0$ corresponds to $u = H^{-1}$. The infinite future is $u \rightarrow
0^+$.

Perturbation theory is organized most conveniently in terms of a 
``pseudo-graviton'' field, $\psi_{\mu \nu}$, obtained by conformally 
re-scaling the metric:
\begin{equation}
g_{\mu \nu} \equiv \Omega^2 {\widetilde g}_{\mu \nu} \equiv 
\Omega^2 \left(\eta_{\mu \nu} + \kappa \psi_{\mu \nu}\right) \; .
\end{equation}
Our notation is that pseudo-graviton indices are raised and lowered with 
the Lorentz metric, and that the loop counting parameter is $\kappa^2 \equiv 
16 \pi G$. After some judicious partial integrations the invariant part of 
the bare Lagrangian takes the following form \cite{tw4}:
\begin{eqnarray}
{\cal L}_{\rm inv} & = & \sqrt{-{\widetilde g}} {\widetilde g}^{\alpha 
\beta} {\widetilde g}^{\rho \sigma} {\widetilde g}^{\mu \nu} 
\left(\frac12 \psi_{\alpha \rho , \mu} \psi_{\nu \sigma , \beta} - 
\frac12 \psi_{\alpha \beta , \rho} \psi_{\sigma \mu , \nu} + \frac14 
\psi_{\alpha \beta , \rho} \psi_{\mu \nu , \sigma} \right. \nonumber \\
& & \mbox{} \left. - \frac14 \psi_{\alpha \rho , \mu} \psi_{\beta \sigma , 
\nu}\right) \Omega^2 -\frac12 \sqrt{-{\widetilde g}} {\widetilde g}^{
\rho \sigma} {\widetilde g}^{\mu \nu} \psi_{\rho \sigma , \mu} 
\psi_{\nu}^{~\alpha} (\Omega^{2})_{,\alpha} \; . \label{eq:Lagrangian}
\end{eqnarray}
Since $\Omega \sim u^{-1}$, it might seem as if the final term is stronger
at late times than the others. In reality it is only comparable because its
undifferentiated pseudo-graviton field must always contain a ``$0$'' index
--- $\psi_{\nu}^{~\alpha} (\Omega^2)_{,\alpha} = 2 u^{-1} \psi_{\nu 0} 
\Omega^2$ --- and ``$0$'' components of the pseudo-graviton propagator are
suppressed by a factor of $u$ \cite{tw3}.

Gauge fixing is accomplished through the addition of $-\frac12 \eta^{\mu \nu} 
F_{\mu} F_{\nu}$ where:
\begin{equation}
F_{\mu} \equiv \left(\psi^{\rho}_{~\mu , \rho} - \frac12 \psi^{\rho}_{~\rho , 
\mu} + 2 \psi^{\rho}_{~\mu} \; {(\ln \Omega)}_{,\rho}\right) \Omega \; .
\end{equation}
The resulting gauge fixed kinetic operator has the form:
\begin{eqnarray}
D_{\mu \nu}^{~~ \rho \sigma} & \equiv & \left(\frac12 {\overline \delta}_{
\mu}^{~(\rho} \; {\overline \delta}_{\nu}^{~\sigma)} - \frac14 \eta_{\mu \nu} 
\; \eta^{\rho\sigma} - \frac12 \delta_{\mu}^{~0} \; \delta_{\nu}^{~0} \; 
\delta_0^{~\rho} \; \delta_0^{~\sigma} \right){\rm D}_A \nonumber \\
& & \mbox{} + \delta_{(\mu}^{~~0} \; {\overline \delta}_ {\nu)}^{~~(\rho} \; 
\delta_0^{~\sigma)} \; {\rm D}_B + \delta_{\mu}^{~0} \; \delta_{\nu}^{~0} \; 
\delta_0^{~\rho} \; \delta_0^{~\sigma} \; {\rm D}_C \; . \label{eq:kinetic}
\end{eqnarray}
A variety of conventions in this relation deserve comment. First, indices 
enclosed in a parenthesis are symmetrized. Second, the presence of a bar over 
a Kronecker delta or a Lorentz metric indicates that the temporal components 
of these tensors are deleted:
\begin{equation}
{\overline \delta}^{\mu}_{~\nu} \equiv \delta^{\mu}_{~\nu} - \delta^{\mu}_{~0}
\delta^0_{~\nu} \qquad , \qquad {\overline \eta}_{\mu\nu} \equiv \eta_{\mu\nu}
+ \delta^0_{~\mu} \delta^0_{~\nu} \; .
\end{equation}
The symbol $D_A$ stands for the kinetic operator of a massless, minimally 
coupled scalar:
\begin{equation}
{\rm D}_A \equiv \Omega \left(\partial^2 + \frac2{u^2}\right) \Omega \; ,
\end{equation}
while $D_B = D_C$ denote the kinetic operator of a conformally coupled scalar:
\begin{equation}
{\rm D}_B = {\rm D}_C \equiv \Omega \> \partial^2 \Omega \; .
\end{equation}

What we actually computed was the amputated expectation value of $\kappa \psi_
{\mu \nu} (u,{\vec x})$ which, on general grounds, must have the following
form:
\begin{equation}
 D_{\mu \nu}^{~~\rho \sigma} \; \left\langle 0 \left\vert \;
\kappa \psi_{\rho \sigma}(x) \; \right\vert 0 \right\rangle =
a(u) \; {\overline \eta}_{\mu \nu} + c(u) \;
\delta^0_{~\mu} \delta^0_{~\nu} \; . \label{eq:amputated}
\end{equation}
Attaching the external leg gives the invariant element, but in a perturbatively
corrected version of conformal coordinates:
\begin{equation}
{\widehat g}_{\mu \nu}(t,{\vec x}) \; dx^{\mu} dx^{\nu} = - \Omega^2 \left[1 
- C(u)\right] \; du^2 + \Omega^2 \left[1 + A(u)\right] \; d{\vec x} \cdot 
d{\vec x} \; . \label{eq:conformal}
\end{equation}
The external leg of the 1-point function is a retarded Green's function in
Schwinger's formalism. From the gauge fixed kinetic operator (\ref{eq:kinetic})
we see that the coefficient functions $A(u)$ and $C(u)$ have the following
expressions in terms of the scalar retarded propagators acting on $a(u)$ and
$c(u)$ \cite{tw5}:
\begin{eqnarray}
A(u) & = & -4 G^{\rm ret}_A[a](u) + G^{\rm ret}_C[3a + c](u) \; ,\label{eq:A}\\
C(u) & = & G^{\rm ret}_C[3a + c](u) \; . \label{eq:C}
\end{eqnarray}

It is simple to work out what the retarded propagators of $D_A$ and $D_C$ 
give when acting on any power of the conformal time:
\begin{eqnarray}
\lefteqn{G^{\rm ret}_A[u^{-4} (Hu)^{\varepsilon}] =  {H^2 \over \varepsilon 
(3 - \varepsilon)} \left\{ (Hu)^{\varepsilon} - 1 + {1 \over 3} \varepsilon -
{1 \over 3} \varepsilon (H u)^3\right\} \; , } \label{eq:GA} \\
\lefteqn{G^{\rm ret}_C[u^{-4} (Hu)^{\varepsilon}] = } \nonumber \\
& & {H^2 \over (1 - \varepsilon) (2 - \epsilon)} \left\{- (Hu)^{\varepsilon} 
+ (2 - \varepsilon) H u - (1 - \varepsilon) (H u)^2\right\} \; . \label{eq:GC}
\end{eqnarray}
One-particle-irreducible diagrams containing $\ell$ loops can be shown to
contribute to $a(u)$ and $c(u)$ at late times no more strongly than some
number times \cite{tw1}:
\begin{equation}
\kappa^{2\ell} H^{2\ell-2} u^{-4} \ln^{\ell}(Hu) \; .
\end{equation}
The action of the scalar retarded propagators on such a term is obtained by
differentiating (\ref{eq:GA}-\ref{eq:GC}) $\ell$ times with respect to 
$\varepsilon$ and then taking the limit $\varepsilon \rightarrow 0$. Because
of the factor of $\varepsilon^{-1}$ on the right hand side of (\ref{eq:GA}),
$G^{\rm ret}_A$ acquires an extra logarithm whereas $G^{\rm ret}_C$ does not.
For example, the leading contributions at two loops give:
\begin{eqnarray}
\lefteqn{G^{\rm ret}_A\left[\kappa^4 H^2 u^{-4} \ln^2(Hu)\right] =} \nonumber \\
& &  (\kappa H)^4 \left\{{1 \over 9} \ln^3(Hu) + {1 \over 9} \ln^2(H u) + 
{2 \over 27} \ln(H u) + {2 \over 81} - {2 \over 81} (H u)^3\right\} \; ,\\
\lefteqn{G^{\rm ret}_C\left[\kappa^4 H^2 u^{-4} \ln^2(Hu)\right] =} \nonumber \\
& & (\kappa H)^4 \left\{-{1 \over 2} \ln^2(H u) - {3 \over 2} \ln(H u) - 
{7 \over 4} + 2 H u - {1 \over 4} (H u)^2\right\} \; .
\end{eqnarray}
This phenomenon has great significance. Its physical origin is the fact that 
$A$-type Green's functions receive contributions from throughout the timelike 
region inside the past lightcone while the $C$-type Green's functions have
support only on the lightlike surface of the past lightcone \cite{tw4}.

Comparison between (\ref{eq:element}) and (\ref{eq:conformal}) results in
the following formulae for the conversion to co-moving time:
\begin{eqnarray}
d(Ht) & = & - \sqrt{1 - C(u)} \;\; d\left[\ln(Hu)\right] \; , \label{eq:con1} \\
b(t) & = & -\ln(Hu) + \frac12 \ln\left[1 + A(u)\right] \; . \label{eq:con2}
\end{eqnarray}
It is then straightforward to work out the relation between physically
interesting quantities defined in co-moving coordinates and the things one
actually computes in perturbation theory. For example, the effective Hubble
constant is:\footnote{Note that the effective Hubble constant is an invariant
by virtue of its relation to the Einstein tensor, $G_{00} = 3 \dot{b}^2$, and
by the fact that co-moving coordinates are unique up to constant rescalings of 
space. It can also be shown to be gauge independent \cite{tw1}.}
\begin{equation}
H_{\rm eff}(t) \equiv {db(t) \over dt} = {H \over \sqrt{1 - C(u)}} \; 
\left\{ 1 - \frac12 u {d \over du} \ln\left[1 + A(u)\right]\right\} \; .
\label{eq:heff}
\end{equation}
Two particularly interesting quantities come from the induced stress tensor:
the energy density $T_{00} = \rho(t)$ and the pressure $T_{ij} = p(t) g_{ij}$. 
The task of this section is completed by first using (\ref{eq:stress}) to
express these in terms of $b(t)$ and then converting to the coefficient 
functions $A(u)$ and $C(u)$:
\begin{eqnarray}
\rho(t) & = & {1 \over 8 \pi G} \left(3 \dot{b}^2(t) - 3 H^2\right) \; , 
\nonumber \\ \label{eq:rho}
& = & {1 \over 8 \pi G} {3 H^2 \over 1 - C} \left\{C - {u A' \over 1 + A} + 
{1 \over 4} \left({u A' \over 1 + A}\right)^2\right\} \; , \\
p(t) & = & - {2 \ddot{b}(t) \over 8 \pi G} - \rho(t) \; , \nonumber \\
\label{eq:pressure}
& = & {1 \over 8 \pi G} {H^2 \over 1 - C} \left\{ {u C' \over 1 - C}
\left[1 - {1 \over 2} {u A' \over 1 + A}\right] - {u \left(u A'\right)' \over 
1 + A} + \left({u A' \over 1 + A}\right)^2\right\} \nonumber \\
& & \mbox{} - \rho(t) \; .
\end{eqnarray}
A dot in these formulae indicates differentiation with respect to $t$, while 
a prime denotes differentiation with respect to $u$.

\section{Two Loop Dominance}

Our perturbative work \cite{tw1} produced explicit results for the late time
($u \rightarrow 0^+$) behavior of the coefficient functions $a(u)$ and $c(u)$ 
at two loops:
\begin{eqnarray}
a(u) & = & H^{-2} \left({\kappa H \over 4 \pi u}\right)^4 \left\{ 
-43 \ln^2(Hu) + {\rm (subleading)}\right\} + O(\kappa^6) \label{eq:a} \\
c(u) & = & H^{-2} \left({\kappa H \over 4 \pi u}\right)^4 \left\{ 
15 \ln^2(Hu) + {\rm (subleading)}\right\} + O(\kappa^6) \; . \label{eq:c}
\end{eqnarray}
We also obtained the following limit on the maximum possible late time
correction to $a(u)$ and $c(u)$ from one-particle-irreducible (1PI) graphs
containing $\ell$ loops:
\begin{equation}
\kappa^{2\ell} H^{2\ell-2} u^{-4} \ln^{\ell}(Hu) \; .
\end{equation}
This bound seemed to suggest that all orders become strong at the same time:
\begin{equation}
-\ln(H u) \sim {1 \over \kappa^2 H^2} = {3 \over 8 \pi} {1 \over G \Lambda}
\gg 1 \; .
\end{equation}
That conclusion is valid for the non-invariant quantities $a(u)$ and $c(u)$,
but not for invariants such as the effective Hubble constant, the induced 
energy density and the induced pressure. The purpose of this section is to
show that, for these quantities, two loop effects become strong at a time
when higher loop corrections are still insignificant. We will also use the
two loop results to derive explicit formulae for the physical invariants
which are valid until perturbation theory breaks down.

The key is the extra logarithm which the spatial trace coefficient $A(u)$
acquires from the $A$-type Green's function. The 1PI amputated coefficient 
functions have the following form:
\begin{eqnarray}
a_{1PI}(u) & = & \sum_{\ell=2}^{\infty} a_{\ell} \kappa^{2\ell} H^{2\ell-2} 
u^{-4} \ln^{\ell}(Hu) + {\rm subdominant} \; , \\
c_{1PI}(u) & = & \sum_{\ell=2}^{\infty} c_{\ell} \kappa^{2\ell} H^{2\ell-2} 
u^{-4} \ln^{\ell}(Hu) + {\rm subdominant} \; ,
\end{eqnarray}
where $a_{\ell}$ and $c_{\ell}$ are pure numbers. The non-amputated coefficient
functions are defined by acting retarded Green's functions according to 
relations (\ref{eq:A}-\ref{eq:C}). From the general action of the retarded
propagators (\ref{eq:GA}-\ref{eq:GC}), we obtain expansions for the leading 
terms induced by 1PI graphs:\footnote{These terms are 1PI except for the
external propagator.}
\begin{eqnarray}
A_{1PI}(u) & = & - {4 \over 3} \ln(H u) \sum_{\ell=2}^{\infty} {a_{\ell} \over
\ell + 1} \left( \kappa^2 H^2 \ln(H u)\right)^{\ell} + {\rm subdominant} \, \\
C_{1PI}(u) & = & - {1 \over 2} \sum_{\ell=2}^{\infty} (3 a_{\ell} + c_{\ell})
\left( \kappa^2 H^2 \ln(H u)\right)^{\ell} + {\rm subdominant} \; .
\end{eqnarray}
From expression (\ref{eq:con2}) for $b(t)$ we see that inflation stops when
$A(u)$ approaches $-1$. The $\ell=2$ term in $A(u)$ passes through $-1$ when:
\begin{equation}
-\ln(H u) = \left({-9 \over 4 a_2}\right)^{\frac13} \left({1 \over \kappa H}
\right)^{\frac43} \; .
\end{equation}
At this time the higher $\ell$ effects in $A(u)$ are of strength:
\begin{equation}
\ln^{\ell+1}(H u) (\kappa H)^{2\ell} \sim (\kappa H)^{\frac23 \ell - \frac43}
\; .
\end{equation}
This is insignificant when one recalls that $\kappa H \sim 10^{-5}$, even for 
GUT scale inflation. And {\it all} the terms in $C(u)$ are insignificant 
because they have one fewer power of the large logarithm.

We have still to account for tadpoles coming from the shift of the background. 
One does this by shifting the fields of the interaction Lagrangian 
(\ref{eq:Lagrangian}) and studying the effect of the induced interactions. For 
example, most 3-point vertices have the generic form: $\psi \partial \psi 
\partial \psi$. Suppose that the two differentiated fields are taken by the
lowest order $A(u)$ terms. This gives a 1-point interaction whose coefficient 
is:
\begin{equation}
\kappa \Omega^2 {d \over du} \left(\kappa^3 H^4 \ln^3(Hu)\right) {d \over du}
\left(\kappa^3 H^4 \ln^3(Hu)\right) \sim \kappa^7 H^6 u^{-4} \ln^4(Hu) \; .
\end{equation}
When one accounts for the extra factor of $\kappa$ in our definition 
(\ref{eq:amputated}) the result is no stronger than the 1PI terms already 
allowed for at $\ell=4$ loops. In fact one can do considerably better at
higher order, but never good enough to catch up with the two loop effect. The
fastest possible growth for either $A(u)$ or $C(u)$ is:
\begin{equation}
(\kappa H)^{4N+8} \ln^{3N+5}(Hu) \; ,
\end{equation}
starting at $N=0$. (The order $(\kappa H)^4$ and $(\kappa H)^6$ terms are 
purely 1PI.) When the two loop term becomes of order one these contributions
are still suppressed by a factor of the small number $(\kappa H)^{4/3} 
\ltwid 10^{-7}$.

It remains to obtain the promised all-orders results for the dominant late
time behavior of $H_{\rm eff}(t)$, $\rho(t)$ and $p(t)$. To simplify the
formulae we define the following small parameter:
\begin{equation}
\epsilon \equiv \left({\kappa H \over 4 \pi}\right)^2 = {G \Lambda \over 3 
\pi} = {8 \over 3} \left({M \over M_P}\right)^4 \; .
\end{equation}
Our explicit two loops results (\ref{eq:a}-\ref{eq:c}) imply:
\begin{eqnarray}
A(u) & = & \epsilon^2 \left\{ {172 \over 9} \ln^3(Hu) + {\rm 
(subleading)}\right\} + O(\epsilon^3) \; , \\
C(u) & = & \epsilon^2 \left\{ 57 \ln^2(Hu) + {\rm (subleading)}\right\} + 
O(\epsilon^3) \; .
\end{eqnarray}
From $C(u)$ and relation (\ref{eq:con1}) we infer the transformation to 
co-moving time:
\begin{equation}
Ht = - \left\{1 - {19 \over 2} \epsilon^2 \ln^2(H u) + \dots \right\} 
\ln(H u) \; .
\end{equation}
This can be inverted to give:
\begin{equation}
\ln(H u) = - \left\{1 + {19 \over 2} (\epsilon H t)^2 + \dots \right\} H t \; .
\end{equation}
It follows that we may set $\ln(H u)$ to $- H t$, to a very good approximation,
for as long as perturbation theory remains valid.

We can now write $A(u)$ as a function of the co-moving time:
\begin{equation}
A(u) = - {172 \over 9} \epsilon^2 (H t)^3 + \dots \; .
\end{equation}
The higher corrections are again insignificant when the first term becomes of
order unity. We can also obtain $b(t)$ as an explicit function of time:
\begin{equation}
b(t) \approx H t + {1 \over 2} \ln(1 + A) \; .
\end{equation}
Substituting into (\ref{eq:heff}) gives the effective Hubble constant:
\begin{eqnarray}
H_{\rm eff}(t) & \approx & H + {1 \over 2} {d \over dt} \ln(1 + A) \; , \\
& \approx & H \left\{1 - {\frac{86}3 \epsilon^2 (H t)^2 \over 1 - \frac{172}9 
\epsilon^2 (H t)^3}\right\} \; . \label{eq:Hexplicit}
\end{eqnarray}
Note that the numerator of the correction term is still quite small when the 
denominator blows up. This is why we are justified in neglecting other terms
--- from $C(u)$ --- which are also of order $(\epsilon H t)^2$.

Going through the same exercise for the induced energy density gives:
\begin{eqnarray}
\rho(t) & \approx & {\Lambda \over 8 \pi G} \left\{- {1 \over H} {\dot{A} \over 
1 + A} + {1 \over 4 H^2} \left({\dot{A} \over 1 + A}\right)^2\right\} \; , \\
& \approx & {\Lambda \over 8 \pi G} \left\{- {\frac{172}3 \epsilon^2 (H t)^2 
\over 1 - \frac{172}9 \epsilon^2 (H t)^3} + \left({\frac{86}3 \epsilon^2 
(H t)^2 \over 1 - \frac{172}9 \epsilon^2 (H t)^3}\right)^2 \right\} \; .
\end{eqnarray}
Note that we cannot neglect the single denominator term compared to the 
double one; in fact the former dominates the latter. The most useful form in
which to give the pressure is added to the energy density:
\begin{eqnarray}
\rho(t) + p(t) & \approx & - {1 \over 8 \pi G} {d^2 \over dt^2} \ln(1 + A) 
\; , \\
& \approx & {1 \over 8 \pi G} \left({ \dot{A} \over 1 + A}
\right)^2 \; , \\
& \approx & {H^2 \over 8 \pi G} \left({\frac{172}3 \epsilon^2 (H t)^2 
\over 1 - \frac{172}9 \epsilon^2 (H t)^3}\right)^2 \; .
\end{eqnarray}
Note that in passing to the middle expression we have neglected the 
term, $\ddot{A}/(1 + A)$, which is still insignificant when $\dot{A}/(1+A)$
is of order one. Note also that when $\dot{A}/(1 + A)$ is small, its square
is even smaller. Therefore $\rho + p$ is quite near zero until screening 
becomes significant.

\section{Discussion}

Our previous work \cite{tw1,tw3,tw4} has established that quantum 
gravitational corrections slow the expansion of an initially inflating 
universe by an amount that becomes non-pertutbatively large at late times.
In this paper we have exploited exact relations between the objects which are 
actually computed in perturbative quantum gravity and the invariant quantities 
of physical interest. We conclude that the mechanism by which perturbation 
theory breaks down is the approach to $-1$ of the spatial trace coefficient 
$A(u)$. Furthermore, this approach is effected by two loop corrections at a 
time well before the higher loop corrections have become significant. 

This insight has a number of consequences, starting with a revised estimate 
for the number of inflationary e-foldings:
\begin{equation}
N \sim \left({9 \over 172}\right)^{\frac13} \left({3 \pi \over G \Lambda}
\right)^{\frac23} = \left({81 \over 11008}\right)^{\frac13} \left({M_P \over
M}\right)^{\frac83} \; ,
\end{equation}
where $M$ is the mass scale of inflation and $M_P$ is the Planck mass. For
inflation on the GUT scale this gives $N \sim 10^7$ e-foldings. Electroweak
inflation should last about $N \sim 10^{45}$ e-foldings. These numbers are 
smaller than our previous estimates, but still much longer than in typical
models. We stress that this long period of inflation is a natural consequence
of the fact that gravity is a weak interaction.

One can also estimate the rapidity with which inflation ends once the effect
becomes noticeable. Suppose we expand around the critical time:
\begin{equation}
H t = N - H {\Delta t} \; .
\end{equation}
When $H {\Delta t} \ll N$ our expression (\ref{eq:Hexplicit}) for the effective
Hubble constant becomes:
\begin{equation}
H_{\rm eff}(t) \approx H \left\{1 - {1 \over 2 H {\Delta t}}\right\} \; .
\end{equation}
It follows that inflation must end rapidly. To be precise, let us define the
end of inflation as the period from when the effective Hubble constant falls
from $\frac9{10}$ to $\frac1{10}$ of its initial value. From the previous
formula, $H_{\rm eff}$ reaches $\frac9{10} H$ at $H {\Delta t} = 5$, and it
falls to $\frac1{10} H$ at $H {\Delta t} = \frac59$, making for a transition
time of $4 \frac49$ e-foldings. 

Of course one cannot trust perturbation theory during this period but it is 
reasonable to conclude that the end of inflation is likely to be sufficiently 
violent to give a substantial amount of re-heating. The end of inflation is 
also likely to be sudden enough to justify assuming that the observationally
relevant density perturbations crossed the causal horizon during the period
when our perturbative expressions are still valid. This means that we do not
need to solve the non-perturbative problem in order to make predictions.

\begin{center}
ACKNOWLEDGEMENTS
\end{center}

We thank the University of Crete for its hospitality during the execution of 
this project. This work was partially supported by DOE contract 
DE-FG02-97ER41029, by NSF grant 94092715, by EEC grant 961206, by NATO 
Collaborative Research Grant 971166 and by the Institute for Fundamental 
Theory.

\end{document}